\documentclass[10pt]{article}
\usepackage{amsfonts}
\usepackage{amssymb}
\usepackage{amsthm}
\usepackage{times}
\newcommand{\states}[1]{\mbox{$\mathcal{P}(\mathfrak{#1})$}}

\newcommand{\abs}[1]{|#1|}
\newcommand{\norm}[1]{\| #1\|}

\newcommand{\alg}[1]{\mbox{$\mathfrak{#1}$}}

\newcommand{\hil}[1]{\mbox{$\mathcal{#1}$}}

\newcommand{\ket}[1]{| #1 \rangle}
\newcommand{\bra}[1]{\langle #1 |}

\newtheorem{lemma}{Lemma}
\newtheorem{thm}{Theorem}
\theoremstyle{definition}
\newtheorem*{defn}{Definition}
\theoremstyle{remark}

\newcommand{\onept}{\mathbb{N}^{*}\!\times \mathbb{N}^{*}}

\begin{document}

\title{Characterizing Quantum Theory in terms of
 Information-Theoretic Constraints}
 \vspace{1in}

\author{\textbf{Rob Clifton\footnote{Rob Clifton died of cancer on
July 31, 2002, while we were working on this project. The final
version of the paper reflects his substantial input to an earlier
draft, and
extensive mutual discussions and correspondence.}} \\
Department of Philosophy,
University of Pittsburgh, \\ Pittsburgh, PA 15260\\
\textbf{Jeffrey Bub} \\
Department of Philosophy, University of Maryland, \\
College Park, MD 20742 \\
(E-mail: jbub@carnap.umd.edu) \\
\textbf{Hans Halvorson} \\
Department of Philosophy, Princeton University, \\
Princeton, NJ 08544 \\
(E-mail: hhalvors@princeton.edu)}
\date{  }

\maketitle
\newpage
 \begin{center}
 ABSTRACT
 \end{center}
 \begin{quote} We show that three fundamental information-theoretic
 constraints---the impossibility of superluminal information transfer
 between two physical systems by performing measurements on one of
them,
 the impossibility of broadcasting the information contained in an
unknown
 physical state, and the impossibility of unconditionally secure bit
 commitment---suffice to
 entail that the observables and state space of a physical theory are
 quantum-mechanical. We demonstrate the converse derivation in part,
 and consider the implications of alternative answers to a remaining
 open question about nonlocality and bit commitment.
 \end{quote}

 \textbf{1. Introduction}

 \textbf{2. The $C^{*}$-Algebraic Approach to Physical Theory}

 \ \ \ \ \textbf{2.1 Basic Concepts}

 \ \ \ \ \textbf{2.2 Physical Generality of The $C^{*}$-Algebraic
Language}

 \ \ \ \ \textbf{2.3 Classical versus Quantum Theories}

 \textbf{3. Technical Results}

 \ \ \ \ \textbf{3.1 Terminology and Assumptions}

 \ \ \ \ \textbf{3.2 No Superluminal Information Transfer via
 Measurement and Kinematic Independence}

 \ \ \ \ \textbf{3.3 No Broadcasting and Noncommutativity}

 \ \ \ \ \textbf{3.3 No Bit Commitment and Nonlocality}

 \textbf{4. Concluding Remarks}

 \newpage

\noindent\footnotesize{Of John Wheeler's `Really Big Questions', the
one on
which most progress has been made is \textit{It from Bit?}---does
information play a significant role at the foundations of physics?
It
is perhaps less ambitious than some of the other Questions, such as
\textit{How Come Existence?}, because it does not necessarily require
a
metaphysical answer.  And unlike, say, \textit{Why the Quantum?}, it
does not require the discovery of new laws of
nature: there was room for hope
that it might be answered through a better
understanding of the laws as we currently know them, particularly
those of quantum physics.  And this is what has happened: the
better understanding is  the quantum theory of information and
computation.

\noindent How might our conception of the quantum physical world have
been
different if \textit{It From Bit} had been a motivation from the
outset?
No one knows how to derive
\textit{it} (the nature of the physical world)  from \textit{bit}
(the idea that
information plays a significant role at the foundations of physics),
and I shall argue that this will never be possible.
But we can do the next best thing: we can start from the qubit.}

\normalsize\begin{flushright}
\emph{Introduction to \textbf{David Deutsch}'s (2003) `It From Qubit'}
\end{flushright}
\section{Introduction}

  Towards the end of the passage above, Deutsch is
  pessimistic about the prospects of deducing the nature of the
  quantum world from the idea that information plays a significant
  role at the foundations of physics.  We propose to
  counter Deutsch's pessimism by
  beginning with the assumption that we live in a world in which
there
  are certain constraints on the acquisition, representation, and
  communication of information, and then deducing from these
assumptions the
  basic outlines of the quantum-theoretic description of
  physical systems.\footnote{Chris Fuchs and Gilles Brassard first
  suggested the project to one of us (JB) as a conjecture or
  speculation (Brassard's preferred term) that quantum mechanics can
  be derived from two cryptographic principles: the possibility of
secure
key distribution and the impossibility of secure bit commitment
\cite{Fuchs1,Fuchs2,Fuchs3,Brassard}.}

  The three fundamental information-theoretic constraints we shall be
  interested in are:
  \begin{itemize}
  \item the
  impossibility of
 superluminal information transfer between two physical systems by
performing
 measurements on one of them;
 \item the impossibility of perfectly
 broadcasting the information contained in an unknown physical
  state; and
 \item the impossibility of unconditionally
 secure bit commitment.
 \end{itemize} These three `no-go's' are all well-known consequences
 of standard nonrelativistic Hilbert space quantum theory.  However,
 like Einstein's radical re-derivation of Lorentz's transformation
based
 upon privileging a few simple principles, we
 here propose to raise the above constraints to the level of
fundamental
 information-theoretic `laws of nature' from which quantum theory
can,
 we claim, be deduced.  We shall do this by starting with a
mathematically
 abstract characterization of a physical theory that includes, as
 special cases, all classical mechanical theories of both wave and
 particle varieties, and all variations on quantum theory, including
 quantum field theories (plus any hybrids of these theories).
 Within this framework,
 we are able to give general formulations of the
  three information-theoretic constraints above, and then show that
they jointly
 entail:
 \begin{itemize}
 \item that the algebras of observables pertaining to
 distinct physical systems must commute, usually called
microcausality
 or (a term we prefer) \emph{kinematic
 independence} (see Summers \cite{Summers});
 \item that any individual
 system's algebra of observables must be nonabelian, i.e.,
 \emph{noncommutative};
 \item that the physical world must be \emph{nonlocal}, in that
 spacelike separated systems must at least sometimes occupy entangled
states.
 \end{itemize}
 We shall argue that these latter three \emph{physical}
characteristics
 are definitive of
 what it means to be
 a quantum theory in the most general sense.  Conversely, we would
 want to
 prove that these three physical characteristics entail
 the three information-theoretic principles from which we started,
 thereby providing a characterization theorem for quantum theory in
 terms of those principles. In this, we are only partly successful,
 because there remains an open question about bit commitment.

  The fact that
  one can characterize quantum theory (modulo the open question) in
  terms of just a few simple information-theoretic principles not
  only goes some way towards
  answering Wheeler's query `Why the Quantum?' (without,
  \emph{pace} Deutsch, the introduction of new laws), but
  lends credence to the idea that an information-theoretic point of
  view is the right perspective to adopt in relation to quantum
theory.
  Notice, in particular, that our derivation
  links information-theoretic principles directly to the
  very features of quantum theory---noncommutativity and
  nonlocality---that are so conceptually problematic from a purely
  physical/mechanical
  point of view.  We therefore
  suggest substituting for the conceptually problematic
  mechanical perspective on quantum theory an
  information-theoretic perspective.  That is, we are suggesting that
  quantum theory be viewed, not as first and foremost a mechanical
theory
  of waves and
  particles (cf. Bohr's infamous dictum, reported in Petersen
\cite{Petersen}:
  `There is no quantum world.'), but as a
  theory about the possibilities and impossibilities of information
  transfer.

We begin, in section \textbf{2}, by laying out the mathematical
framework within which our entire analysis will be conducted---the
theory of $C^{*}$-algebras.  After introducing the basics, we
establish that the $C^{*}$-algebraic framework does indeed encompass
both
classical and
quantum statistical theories, and go on to argue that the latter
class
of theories is most properly viewed as picked out solely
in virtue of its satisfaction of kinematic independence,
noncommutativity, and nonlocality---even though there is far more
to the physical content of any given quantum theory than these
three features.  Section \textbf{3} contains our technical results.
We
first formulate the three information-theoretic
constraints---no superluminal information transfer
via measurement, no broadcasting, and nobit
commitment---in $C^{*}$-algebraic terms, after briefly reviewing the
concepts as
they occur in standard nonrelativistic Hilbert space quantum theory.
We show that these information-theoretic `no-go' principles jointly
entail
kinematic independence, noncommutativity,
and nonlocality. We
demonstrate the converse derivation in part: that the physical
properties of
kinematic independence and noncommutativity jointly entail
no superluminal information transfer via measurement and
no broadcasting.

The remaining open question concerns the derivation
of the impossibility of unconditionally secure bit commitment from
kinematic independence, noncommutativity, and nonlocality
in the theory-neutral $C^{*}$-algebraic framework. The proof of this
result in standard
quantum mechanics (Mayers \cite{Mayers1,Mayers2}, Lo and Chau
\cite{Lo})
depends on the biorthogonal decomposition
theorem, which is not available in the more general framework. If the
derivation
goes through, then we have a characterization theorem for quantum
theory in terms of the three information-theoretic principles,
thereby
considerably generalizing the known proofs of these principles within
the
standard nonrelativistic Hilbert space quantum theory framework. If
not, then there must be quantum mechanical systems---perhaps systems
associated with von Neumann algebras of some nonstandard type---that
allow an
unconditionally secure bit commitment protocol! Section
\textbf{4} concludes with some further remarks about the significance
of our information-theoretic characterization of quantum theory.

\section{The $C^{*}$-Algebraic Approach to Physical Theory}

\subsection{Basic Concepts}

 We start with a brief review of abstract $C^{*}$-algebras and their
 relation to the standard formulation of quantum theory in terms of
 concrete algebras of operators acting on a Hilbert space.

 A unital
$C^{*}$-\emph{algebra} is a Banach $^{*}$-algebra over $\mathbb{C}$
containing the identity, where the
involution and norm are related by $\|A^{*}A\|=\|A\|^{2}$.  Thus,
the algebra $\alg{B}(\hil{H})$ of all bounded operators on a
Hilbert space
$\hil{H}$---which, of course, is used in the standard formulation of
nonrelativistic
quantum theory---is an example of a $C^{*}$-algebra,
with $^{*}$ taken to be
the adjoint operation, and $\|\cdot\|$ the
standard operator norm.
Moreover, any $^{*}$-\emph{sub}algebra of $\alg{B}(\hil{H})$
containing the identity operator
that is closed in the operator norm is a (unital) $C^{*}$-algebra.
By a
\emph{representation} of a $C^{*}$-algebra $\alg{A}$ is meant any
mapping
$\pi:\alg{A}\rightarrow\alg{B}(\hil{H})$ that preserves the linear,
product, and $^{*}$ structure of \alg{A}.  If, in addition, $\pi$ is
one-to-one (equivalently, $\pi(A) = 0$ implies $A = 0$),
the representation is called \emph{faithful}. In a faithful
representation, $\pi(\alg{A})$ provides an isomorphic copy of
$\alg{A}$. A representation is irreducible just in case the only
closed
subspaces of $\hil{H}$ that are invariant under $\pi$ are $\hil{H}$
and the null space.

A von Neumann algebra $\mathcal{R}$ is a concrete collection of
operators on some
fixed Hilbert space $\hil{H}$---specifically, a $^{*}$-subalgebra of
$\alg{B}(\hil{H})$ that contains the identity and satisfies
$\mathcal{R} = \mathcal{R''}$. (Here $\mathcal{R''}$ is the double
commutant of $\mathcal{R}$, where the  commutant $\mathcal{R'}$
 is the set of all
operators on $\hil{H}$ that commute with every operator in
$\mathcal{R}$).
This is equivalent, via von Neumann's
double commutant theorem (\cite[Theorem 5.3.1]{Kadison}), to the
assertion that $\mathcal{R}$ contains the identity and is closed in
the strong operator topology (where $Z_{n} \rightarrow Z$ strongly
just
in case $\| (Z_{n} - Z)x \| \rightarrow 0$ for all $x
\in \hil{H}$, and the norm here is the Hilbert space vector norm).

Every von Neumann algebra is also a $C^{*}$-algebra, but not every
$C^{*}$-algebra of operators is a von Neumann algebra. A von Neumann
algebra $\mathcal{R}$ is termed a factor just in case its center
$\mathcal{R} \cap \mathcal{R'}$ contains only multiples of the
identity. This is equivalent to the condition
$(\mathcal{R} \cup \mathcal{R'})'' = \alg{B}(\hil{H})$, so
$\mathcal{R}$ induces a `factorization' of the total Hilbert space
algebra $\alg{B}(\hil{H})$ into two subalgebras which together
generate that algebra. Factors are classified into different types.
The algebra $\alg{B}(\hil{H})$ for any Hilbert space $\hil{H}$ is a
type I factor, and every type I factor arises as the algebra of all
bounded operators on some Hilbert space
\cite[Theorem 6.6.1]{Kadison}.
Type II and type III factors, and subclassifications, have
applications
to
the thermodynamic limit of quantum statistical mechanics and
quantum field theory.

A \emph{state} of a $C^{*}$-algebra $\alg{A}$ is taken to be
any positive, normalized,
linear functional $\rho:\alg{A}\rightarrow\mathbb{C}$ on the
algebra.  For
example, a state of $\alg{B}(\hil{H})$ in standard quantum theory is
obtained
if we select
a positive trace-one (density) operator $D$ on $\hil{H}$, and define
$\rho(A)=\mbox{Tr}(AD)$ for all $A\in \alg{B}(\hil{H})$, which
defines a
linear functional that is
positive (if $A=X^{*}X$, then $\mbox{Tr}(X^{*}XD)=
\mbox{Tr}(XDX^{*})$, and the latter is nonnegative
 because $XDX^{*}$ is a positive operator), normalized (since
 $\mbox{Tr}(ID)=1$), and linear (since operator composition and the
 trace operation are both linear).  We can make the usual distinction
between
 pure and
mixed states, the former defined by the property that if
$\rho=\lambda\rho_{1}+(1-\lambda)\rho_{2}$, with $\lambda\in (0,1)$,
then
$\rho=\rho_{1}=\rho_{2}$.   In the concrete case of
$\alg{B}(\hil{H})$,
a pure state of course corresponds to a density operator for which
$D^{2}=D$---which is equivalent to the existence of a unit vector
$\ket{v}\in\hil{H}$ representing the state of the system via
$\rho(A)=\bra{v}A\ket{v}$ ($A\in \alg{B}(\hil{H})$).

  One should note, however, that, because countable additivity is not
  presupposed by the $C^{*}$-algebraic notion of state (and,
  therefore, Gleason's theorem does not apply),
  there can be pure states of $\alg{B}(\hil{H})$ not
 representable by vectors in $\hil{H}$.
 In fact,
 if $A$ is any self-adjoint element of a
$C^{*}$-algebra $\alg{A}$, and $a\in\mbox{sp}(A)$,
then there always exists a pure state $\rho$ of
$\alg{A}$ that assigns a \emph{dispersion-free} value of $a$ to $A$
\cite[Ex. 4.6.31]{Kadison}.
Since this is true even when we
consider a
point in the
continuous spectrum of a self-adjoint operator $A$ acting on a
Hilbert
space, \emph{without} any corresponding eigenvector,
it follows that there \emph{are} pure states of $\alg{B}(\hil{H})$ in
the $C^{*}$-algebraic sense
that cannot be vector states (nor, in fact, representable by any
density
operator
$\hil{H}$).

The primary reason countable additivity is not required of
 $C^{*}$-algebraic states is that it is a representation-dependent
 concept that presupposes the availability of
 the notion of an infinite convergent sum of
 orthogonal projections.  That is: to say that a state $\rho$ is
 countably additive is to say that
 $\sum_{i=1}^{\infty}P_{i}=I$ implies
 $\sum_{i=1}^{\infty}\rho(P_{i})=1$; yet the notion of convergence
 required for the first sum to make sense is strong operator
 convergence (where a sequence of operators $A_{i}$
 converges strongly to some operator $A$
 just in case, for all $x\in\hil{H}$, $A_{i}x\rightarrow Ax$ in
 \emph{vector} norm). Since, obviously, the elements of a
 $C^{*}$-algebra $\alg{A}$ need not
 be thought of as operators acting on a Hilbert space of vectors,
 countable additivity is simply unavailable as a potential general
 constraint that could be imposed on the state space of $\alg{A}$.

Nevertheless, it turns out that for \emph{any} state $\rho$ of a
$C^{*}$-algebra
$\alg{A}$, there is always \emph{some} representation of $\alg{A}$ in
which $\rho$ \emph{is} representable by a vector (even if $\rho$ is a
mixed state).   According to the
Gelfand-Naimark-Segal theorem \cite[Thm. 4.5.2]{Kadison},
every state $\rho$ determines a unique
(up to unitary equivalence)
representation $(\pi _{\rho},\hil{H}_{\rho})$ of $\alg{A}$ and vector
$\Omega _{\rho}\in \hil{H}_{\rho}$ such that
$\rho (A)=\langle \Omega _{\rho},\pi _{\rho}(A)\Omega _{\rho} \rangle$
($A\in \alg{A}$),
and such that the set $\{ \pi _{\rho}(A)\Omega _{\rho}:A\in \alg{A}\}$
is dense in $\hil{H}_{\rho}$. The triple $(\pi
_{\rho},\hil{H}_{\rho},\Omega
_{\rho})$ is called the GNS representation of $\alg{A}$
induced by the state $\rho$, and this representation is
irreducible if and only if $\rho$ is pure (equivalently,
 every bounded operator on $\hil{H}_{\rho}$ is the
strong  limit
of operators in  $\pi _{\rho}(\alg{A})$).

Now, by considering the
collection of all pure states on $\alg{A}$, and forming the direct
sum
of all the irreducible GNS representations these states determine,
one obtains
a highly reducible but faithful representation of $\alg{A}$ in which
 every pure state of $\alg{A}$ is represented by a vector (as usual
in
 standard nonrelativistic quantum theory).
As a consequence of this construction, we obtain the Gelfand-Naimark
theorem:
every abstract $C^{*}$-algebra has a concrete
faithful representation as a norm-closed $^{*}$-subalgebra of
$\alg{B}(\hil{H})$, for some appropriate Hilbert space $\hil{H}$
\cite[Remark 4.5.7]{Kadison}.
So there is a sense in which
$C^{*}$-algebras are no more general than
algebras of operators on Hilbert spaces---apart from the fact that,
 when working
with an abstract $C^{*}$-algebra, one does not
privilege
any particular concrete Hilbert space representation of the algebra
(which turns out to be important not to do in quantum field theory,
where one needs to allow for inequivalent representations of the
canonical commutation relations---see, e.g., Clifton and Halvorson
 \cite{CliftonHalv2}).

\subsection{Physical Generality of the $C^{*}$-Algebraic Language}

If $C^{*}$-algebras supply little more than a way of talking
abstractly about operator algebras, and the latter are characteristic
of quantum theory, how can we possibly claim that the $C^{*}$-algebraic
machinery
supplies a
universal language within which all mainline physical theories,
including even \emph{classical} mechanics, can be framed?
The fallacy, here, is that the use of operator algebras is
only relevant to quantum theories.

Take, as a simple example, the
classical description of a system of $n$ point
particles. Focusing first on the kinematical content of the theory,
the observables of the system are
real-valued
functions on its phase space $\mathbb{R}^{6n}$.  These can be thought
of as the
self-adjoint elements of the $C^{*}$-algebra
$\alg{B}(\mathbb{R}^{6n})$ of
all bounded, complex-valued
measurable functions on $\mathbb{R}^{6n}$---where the multiplication
law is
just pointwise multiplication of functions, the adjoint is complex
conjugation, and the norm of a function is the supremum of its
absolute values.
The statistical states of the system are given by probability
measures $\mu$ on
$\mathbb{R}^{6n}$, and pure states, corresponding to maximally
complete
information about the particles, are given by
the individual points of
$\mathbb{R}^{6n}$.  Using a statistical state $\mu$, we obtain the
corresponding expectation functional, which is the system's state in
the $C^{*}$-algebraic sense, by defining
$\rho(f)=\int_{\mathbb{R}^{6n}}f\mbox{d}\mu$
($f\in\alg{B}(\mathbb{R}^{6n})$).

Turning now to dynamics, the Heisenberg picture of the
time evolution of a state is
determined by a group of bijective, Lebesgue measure-preserving,
 flow mappings
$T_{t}:\mathbb{R}^{6n}\rightarrow \mathbb{R}^{6n}$ ($t\in\mathbb{R}$)
that induce an automorphism group $\tau_{t}$ on
$\alg{B}(\mathbb{R}^{6n})$
via $\tau_{t}(f)=f\circ T_{t}$.
State evolution when a measurement occurs is
also fully analogous to the quantum case.  The probability in state
$\mu$
that the value of $f$ will be found on measurement to lie
 in a Borel set $\Delta$ is given by
$\rho(\chi_{f^{-1}(\Delta)})$ (note that $\chi_{f^{-1}(\Delta)}$
is a projection in  $\alg{B}(\mathbb{R}^{6n})$); and, should that be
the case,
the new post-measurement state is given by (see B\'{o}na [2000] and
Duvenhage [2002a,b]):
 \[\rho'(g)=
\frac{\rho(\chi_{f^{-1}(\Delta)}g\chi_{f^{-1}(\Delta)})}
{\rho(\chi_{f^{-1}(\Delta)})} \qquad \qquad (g\in\mathbb{R}^{6n})\]

Lastly, note
that because classical $n$ point particle mechanics employs a
$C^{*}$-algebra (as we have seen, $\alg{B}(\mathbb{R}^{6n})$), it
follows from the
Gelfand-Naimark theorem that classical mechanics can be done in
Hilbert space!  Yet this really  is nothing new, having been
pointed out long ago by Koopman \cite{Koopman} and von Neumann
\cite{VonNeumann}
(see Mauro \cite{Mauro} for an up-to-date discussion).

Of course, nothing we have said \emph{proves} that all
physical theories admit a $C^{*}$-algebraic
formulation.  Indeed, that would be absurd to claim: one can
certainly conceive of theories whose algebra of observables falls
short of being isomorphic to the self-adjoint part of a
$C^{*}$-algebra and, instead, only instantiates some weaker
mathematical structure, such  as a Segal \cite{Segal} algebra.  To
foreclose
such possibilities, it could be of interest to pursue
an axiomatic justification of the $C^{*}$-algebraic framework along
lines similar to those provided by
Emch
\cite[Ch. 1.2]{Emch}.   However, it suffices for present purposes
simply to
observe that all physical theories that have  been found
\emph{empirically
successful}---not just phase space and Hilbert space theories
(Landsman \cite{Landsman}), but also theories based a
manifold (Connes \cite{Connes})---fall under this framework (whereas,
for example,
so-called `nondistributive'
Segal algebras permit violations of the Bell inequality far in excess
of that permitted by standard quantum theory and
observed in the laboratory---see Landau \cite{Landau2}).

\subsection{Classical versus Quantum Theories}

We must mention one final important representation theorem: every
(unital)
\emph{abelian} $C^{*}$-algebra $\alg{A}$ is isomorphic to the set
$C(X)$ of
all continuous, complex-valued functions on a compact Hausdorff space
$X$
\cite[Thm. 4.4.3]{Kadison}.   This is called
the
\emph{function representation} of $\alg{A}$.
The underlying `phase space' $X$ in this representation is none other
than the pure state space
$\states{A}$ of
$\alg{A}$ endowed with its weak-* topology. (A sequence of states
$\{\rho_{n}\}$ on $\alg{A}$ weak-* converges to a state $\rho$ just
in
case $\rho_{n}(A) \rightarrow \rho(A)$ for all $A\in\alg{A}$.)
The isomorphism
maps
an element $A\in\alg{A}$ to the function $\hat{A}$ (the Gelfand
transformation of $A$) whose value at any
$\rho\in\states{A}$ is just the (dispersion-free) value that
$\rho$ assigns to $A$.   Thus, not only does every classical phase
space
presentation of a physical theory define a $C^{*}$-algebra, but,
conversely, behind every abstract abelian $C^{*}$-algebra lurks in its
function
representation a good old-fashioned classical
phase space theory.  All of this justifies
treating a theory formulated in $C^{*}$-algebraic language as classical
just in case its algebra is abelian.
It follows that a necessary condition for thinking of a theory as a
\emph{quantum}
theory is that its $C^{*}$-algebra be non-abelian.  However, as we shall
now explain, we do not believe
this is sufficient unless something further is said about the
presence
of entangled states.

In 1935 and 1936, Schr\"{o}dinger published
 an extended two-part
commentary \cite{Schr1,Schr2} on the
Einstein-Podolsky-Rosen argument \cite{Einstein2},
where he introduced the term `entanglement' to describe the peculiar
correlations of the EPR-state as
\cite[p. 555]{Schr1}: `\textit{the} characteristic trait of
quantum mechanics, the one that enforces its entire departure from
classical lines of thought.' In the
first part, he considers entangled states for which the biorthogonal
decomposition is unique, as well as cases like the EPR-state, where
the
biorthogonal decomposition is non-unique. There he is concerned to
show that suitable measurements on one system can fix the (pure)
state of the
entangled distant system, and that this state depends on what
observable one chooses to measure, not merely on the outcome of that
measurement. In the second part, he shows that a
`sophisticated experimenter,' by performing a suitable local
measurement on
one system, can `steer' the
distant system into any mixture of pure states representable by
its reduced density
operator. (So the distant
system can be steered into any pure state in the support of the
reduced density operator, with a nonzero probability that depends
only
on the pure state.) For a mixture of linearly
independent states, the steering can be done by performing a
PV-measurement in a
suitable basis. If the states are linearly dependent, the
experimenter
performs a POV-measurement, which amounts to enlarging the
experimenter's Hilbert space by adding an ancilla, so that the
dimension of the enlarged Hilbert space is equal to the number of
linearly independent states. Schr\"{o}dinger's result here
anticipates the later result by Hughston, Jozsa, and Wootters
 \cite{Hughston} that underlies the
`no go' bit commitment theorem. (Similar results were proved by
Jaynes \cite{Jaynes} and Gisin \cite{Gisin}.)

What Schr\"{o}dinger found problematic---indeed,
objectionable---about
entanglement was this possibility of remote steering
\cite[p. 556]{Schr1}:
\begin{quote}
It is rather discomforting that the theory should allow a system to
be steered or piloted into one or the other type of state at the
experimenter's mercy in spite of his having no access to it.
\end{quote}

He conjectured that an entangled state of a composite system would
almost instantaneously decay to a mixture as the component systems
separated. (A similar possibility was raised and rejected by Furry
\cite{Furry}.)
There would still be correlations between the states of the
component systems, but remote steering would no longer be possible
\cite[p. 451]{Schr2}:
\begin{quote}
    It seems worth noticing that the [EPR] paradox could be avoided
by a
    very simple assumption, namely if the situation after separating
    were described by the expansion (12), but with the additional
    statement that the knowledge of the \textit{phase relations}
    between the complex constants $a_{k}$ has been entirely lost in
    consequence of the process of separation. This would mean that
    not only the parts, but the whole system, would be in the
    situation of a mixture, not of a pure state. It would not preclude
    the possibility of determining the state of the first system by
    \textit{suitable} measurements in the second one or \textit{vice
    versa}. But it would utterly eliminate the experimenters influence
    on the state of that system which he does not touch.
\end{quote}

\noindent Expansion (12) is the biorthogonal expansion:
\noindent
\begin{equation}
\Psi(x,y) = \sum_{k}a_{k}g_{k}(x)f_{k}(y)
\end{equation}

Schr\"{o}dinger regarded the phenomenon of
interference associated with noncommutativity
in quantum mechanics as unproblematic, because he saw this as
reflecting the fact that particles are
wavelike.
But he did not believe that we live
in a world in which physical systems can exist nonlocally in
entangled
states, because such states would allow Alice to steer Bob's system into any
mixture of pure states compatible with Bob's reduced density operator.
Schr\"{o}dinger did not expect that
experiments would bear this out. On his view, entangled
states, which the theory allows, are entirely local
insofar as they characterize physical systems, and
nonlocal entangled states are simply an artefact of the formalism.

Of course, it was an experimental question in 1935 whether
Schr\"{o}dinger's conjecture was correct or not.
We now know that the conjecture is false. A
wealth of experimental evidence, including the experimentally
confirmed violations of Bell's inequality (e.g., Aspect \textit{et al}
\cite{Aspect}),
testify to this. The
relevance of Schr\"{o}dinger's conjecture for our inquiry is this: it
raises the possibility of a quantum-like world in which there is
interference but
no nonlocal entanglement. We will need to exclude this possibility on
information-theoretic grounds.

As indicated, for a composite system, A+B, consisting of two
component subsystems, A and B, we propose to show (i) that the
`no superluminal information transfer via measurement'
condition entails that the
$C^{*}$-algebras $\alg{A}$ and $\alg{B}$, whose self-adjoint elements
represent the observables of A and B,
commute with each other, and (ii) that the `no broadcasting'
condition
entails that $\alg{A}$ and $\alg{B}$ separately are noncommutative
(nonabelian).
Now, if $\alg{A}$ and $\alg{B}$ are nonabelian and mutually commuting
(and $C^{*}$-independent\footnote{See section 3.1.}), it follows
immediately that there are nonlocal entangled states on the
$C^{*}$-algebra
$\alg{A}\vee\alg{B}$ they generate (see Landau \cite{Landau1}, who
shows that there is a state $\rho$ on $\alg{A}\vee\alg{B}$ that
violates Bell's inequality and hence is nonlocally entangled; also
Summers and Werner \cite{Summers} and
Bacciagaluppi \cite{Bacc}). So, at least mathematically, the presence
of nonlocal entangled states in the formalism is guaranteed, once we
know that
the algebras of observables are nonabelian. What does not follow
is that these states actually occur in nature. For example, even
though Hilbert space quantum mechanics allows for paraparticle
states, such states are not observed in nature. In terms of our
program, in order to show that
entangled states are actually instantiated,
and---contra Schr\"{o}dinger---instantiated nonlocally,
we need to derive this from some
information-theoretic principle. This is the role of the `no bit
commitment' constraint.

Bit commitment is a cryptographic protocol in which one party, Alice,
supplies an encoded bit
to a second party, Bob. The information available in the encoding
should be insufficient for Bob to ascertain the value of the bit, but
sufficient, together with further information supplied by Alice at a
subsequent stage when she is supposed to reveal the
value of the bit, for Bob to be convinced that the protocol does not
allow Alice to cheat by encoding the bit in a way that leaves her
free
to reveal either 0 or 1 at will.

In 1984, Bennett and Brassard \cite{Bennett} proposed a quantum bit
commitment protocol now referred to as BB84. The basic idea was to
associate
the 0 and 1
commitments with two equivalent quantum mechanical mixtures
represented by the same density operator. As they showed,
Alice can cheat by adopting an Einstein-Podolsky-Rosen
(EPR)
attack or cheating strategy: she prepares entangled pairs of
particles, keeps one of each pair (the ancilla) and sends the second
particle (the channel particle) to Bob. In this way she can fake
sending one of two equivalent mixtures to Bob,
and reveal either bit at will at the opening stage, by
effectively steering Bob's particles into
the desired mixture via appropriate measurements on her ancillas. Bob
cannot detect this cheating strategy.

Mayers \cite{Mayers1,Mayers2}, and Lo and Chau \cite{Lo}, showed that
the insight of
Bennett and Brassard
can be extended to a proof that a generalized version of
the EPR cheating strategy can always be applied, if the
Hilbert space is enlarged in a suitable way by introducing additional
ancilla particles. The proof of the `no go' quantum bit commitment
theorem
exploits biorthogonal decomposition
via the Hughston-Jozsa-Wootters result \cite{Hughston}
(effectively anticipated by
Schr\"{o}dinger). Informally, this  says that for a quantum
mechanical system
consisting of two (separated) subsystems represented by
 the tensor product of two
type-I factors
$\alg{B}(\hil{H}_{1}) \otimes \alg{B}(\hil{H}_{2})$, any mixture of
states on $\alg{B}(\hil{H}_{2})$ can be generated
from a distance by performing an appropriate POV-measurement on the
system represented by $\alg{B}(\hil{H}_{1})$, for an appropriate
entangled state of the composite system
 $\alg{B}(\hil{H}_{1}) \otimes \alg{B}(\hil{H}_{2})$.
This is what makes it possible for Alice to cheat in her bit
commitment protocol with Bob. It is easy enough to see this for the
original BB84 protocol. Suprisingly, this is also the case for
 any conceivable quantum bit commitment
protocol. (See Bub \cite{Bub} for a discussion.)

Now, unconditionally secure bit commitment is impossible for
classical
systems, in which the algebras of observables are abelian.
It might seem inappropriate, then, that we propose `no bit commitment'
as a constraint
distinguishing quantum from classical theories. The relevant point to
note here is that the insecurity of any bit commitment protocol
in a nonabelian setting depends on considerations entirely
different from those in a classical abelian setting. Classically,
unconditionally secure bit commitment is impossible, essentially
because
 Alice can send (encrypted) information to Bob that guarantees the
 truth of an exclusive classical disjunction (equivalent
to her commitment to a 0 or a 1) only if the information is biased
towards one
of the alternative disjuncts (because a classical exclusive
disjunction is true
if and only if one of the disjuncts is true and the other false). No
principle of classical mechanics precludes Bob from extracting this
information.
So the security of the protocol cannot be unconditional and
 can only depend on issues of
computational complexity.

 By contrast, in a
situation of the sort envisaged by Schr\"{o}dinger, in which the
algebras of observables are nonabelian but composite physical systems
cannot exist in nonlocal entangled states,
 if Alice sends Bob one of two mixtures associated with the same
 density operator to establish her commitment, then she
is, in effect, sending Bob evidence for the
truth of an exclusive disjunction that is not based on the selection
of a particular disjunct. (Bob's reduced density operator
is associated ambiguously with both mixtures, and
hence with the truth of the exclusive disjunction: `0 or 1'.) This is
what
noncommutativity allows: different mixtures can be associated with
the same density operator. What thwarts the possibility of using the
ambiguity of mixtures in this way to implement an unconditionally
secure bit commitment protocol is the existence of nonlocal entangled
states between Alice and Bob. This allows Alice to cheat by preparing
a suitable entangled state instead of one of the mixtures, where the
reduced density operator for Bob is the same as that of the mixture.
Alice is
then able to steer Bob's systems into either of the two mixtures
associated with the alternative commitments at will.

So what would allow unconditionally secure bit commitment in a
nonabelian theory is the absence of physically occupied
 nonlocal entangled states.
One can therefore take Schr\"{o}dinger's remarks as relevant to the
question of whether or not secure bit
commitment is possible in our world. In effect, Schr\"{o}dinger
believes
that we live in a quantum-like world in which  secure bit commitment
is possible.
Experiments such as those designed to test Bell's inequality can be
understood as demonstrating that this is not the case. The violation
of Bell's inequalities can then be seen as a criterion for the
possibility of
remote steering in Schr\"{o}dinger's sense.

\section{Technical Results}

\subsection{Terminology and Assumptions}

Our aim in this section is to show that the kinematic aspects of the
quantum theory of a composite system A+B (consisting of two component
subsystems A and B) can be characterized in terms of
information-theoretic constraints.  The physical observables of A and
B are represented, respectively, by self-adjoint elements of unital
subalgebras $\alg{A}$ and $\alg{B}$ of a $C^{*}$-algebra $\alg{C}$.
We let $\alg{A}\vee \alg{B}$ denote the $C^{*}$-algebra generated by
$\alg{A}$ and $\alg{B}$.  A state of A, B, or A+B (i.e., a catalog of
the expectation values of all observables) can be represented by means
of a positive, normalized, linear functional on the respective algebra
of observables.  Recall that a state $\rho$ is said to be pure just in
case $\rho=\lambda\rho_{1}+(1-\lambda)\rho_{2}$, for $\lambda\in
(0,1)$, entails that $\rho=\rho_{1}=\rho_{2}$; otherwise, $\rho$ is
said to be mixed.  For simplicity, we will assume that the systems A
and B are identically constituted --- i.e., they have precisely the
same degrees of freedom --- and therefore that there is an isomorphism
between the algebras $\alg{A}$ and $\alg{B}$.  (However, it will be
clear that most of our results do not depend on this assumption.)  We
will hold this isomorphism fixed throughout our discussion so that we
can use the same notation to denote ambiguously an operator in
$\alg{A}$ and its counterpart in $\alg{B}$.  We will also use the same
notation for a state of $\alg{A}$ and its counterpart in the state
space of $\alg{B}$.

The most general dynamical evolution of a system is represented by a
completely positive, linear `operation' mapping $T$ of the
corresponding algebra of observables.  (We also require that $T(I)\leq
I$.  The operation $T$ is said to be selective if $T(I)<I$, and
nonselective if $T(I)=I$.)  Recall that a linear mapping $T$ of
$\alg{A}$ is positive just in case $A\geq 0$ entails $T(A)\geq 0$, and
is completely positive just in case, for each positive integer $n$,
the mapping $T\otimes \iota$ of $\alg{A}\otimes M_{n}(\mathbb{C})$
into itself, defined by $(T\otimes \iota )(A\otimes B)=T(A)\otimes B$,
is positive.  (Here $M_{n}(\mathbb{C})$ is the $C^{*}$-algebra of
$n\times n$ matrices over the complex numbers.)  If $T$ is an
operation of $\alg{A}$, and $\rho$ is a state of $\alg{A}$ such that
$\rho (T(I))\neq 0$, then the mapping $T^{*}\rho$ defined by
\begin{eqnarray} (T^{*}\rho )(A) &=& \frac{\rho (T(A))}{\rho (T(I))}
\qquad \qquad
(A \in \alg{A}) \end{eqnarray}
is a state of $\alg{A}$.  The standard example of a selective
operation is a collapsing von Neumann
measurement of some observable $O$ with spectral projection $P$.  In
that case, $T(A)=PAP$ ($A\in\alg{A}$), and $T^{*}\rho$ is the final
state obtained after measuring $O$ in state $\rho$ and ignoring all
elements of the ensemble that do not yield as measurement result the
eigenvalue of $O$ corresponding to $P$.  The standard example of a
nonselective operation is a time evolution induced by a unitary
operator $U\in\alg{A}$, where $T(A)=U^{*}AU$ ($A\in\alg{A}$) simply
represents the Heisenberg picture of such evolution.

Finally, we must add one nontrivial independence assumption in order
to capture the idea that A and B are physically \emph{distinct}
systems.  (Our current assumptions would allow that $\alg{A}=\alg{B}$,
which obviously fails to capture the situation we are intending to
describe.)  Various notions of independence for a pair
$\alg{A}$, $\alg{B}$ of $C^{*}$-algebras have been developed in the
literature \cite{Florig,Summers}.  We are particularly interested in
the notion of $C^{*}$-independence developed in \cite{Florig}, because
it does not presuppose that $\alg{A}$ and $\alg{B}$ are kinematically
independent (i.e., that $[A,B]=0$ for all $A\in \alg{A}$ and $B\in
\alg{B}$).  Thus, we will assume that --- whether or not $\alg{A}$ and
$\alg{B}$ are kinematically independent --- any state of $\alg{A}$ is
compatible with any state of $\alg{B}$.  More precisely, for any state
$\rho _{1}$ of $\alg{A}$, and for any state $\rho _{2}$ of $\alg{B}$,
there is a state $\rho$ of $\alg{A}\vee \alg{B}$ such that $\rho
|_{\alg{A}}=\rho _{1}$ and $\rho |_{\alg{B}}=\rho _{2}$.  This
condition holds (i.e., $\alg{A}$ and $\alg{B}$ are
$C^{*}$-independent) if and only if $\norm{AB}=\norm{A}\norm{B}$, for
all $A\in \alg{A}$ and $B\in \alg{B}$ \cite[Prop.~3]{Florig}.

\subsection{No Superluminal Information Transfer via Measurement and Kinematic
Independence}

We first show that $\alg{A}$ and $\alg{B}$ are kinematically
independent if and only if the `no superluminal information transfer via
measurement' constraint holds.  The sense of this constraint is that
 when Alice and Bob perform local measurements,
Alice's measurements can have no influence on the
statistics for the outcomes of Bob's measurements (and vice versa);
for, otherwise, measurement would
allow the instantaneous transfer of information between Alice and Bob.
That is, the mere performance of a local measurement (in the
nonselective sense) cannot, in and of itself, transfer information to
a physically distinct system.

The most
  general nonselective measurement operation performable by Alice is
given~by
\begin{eqnarray} T(A) &=& \sum _{i=1}^{n}E_{i}^{1/2}AE_{i}^{1/2}
  \qquad \qquad (A\in \alg{A}\vee \alg{B}) \end{eqnarray}
where the $E_{i}$ are positive operators in $\alg{A}$ such that $\sum
_{i=1}^{n}E_{i}=I$. The restriction to nonselective measurements is
justified here because selective operations can trivially
change the statistics of observables measured at a distance, simply in
virtue of the fact that the ensemble relative to which one computes
statistics has changed.

We will say that an operation $T$
conveys no information to Bob just in case $T^{*}$ leaves the state of
Bob's system invariant (so that everything `looks the same' to Bob
after the operation as before, in terms of his expectation values for
the outcomes of measurements on observables).
\begin{defn} An operation $T$ on $\alg{A}\vee \alg{B}$ conveys no
information to
  Bob just in case $(T^{*}\rho )|_{\alg{B}}=\rho |_{\alg{B}}$ for all
  states $\rho$ of $\alg{B}$.  \end{defn}

\noindent Note that each $C^{*}$-algebra has sufficient states to discriminate
between any two
observables (i.e., if $\rho (A)=\rho (B)$ for all states $\rho$, then
$A=B$). Now $(T^{*}\rho )|_{\alg{B}}=\rho |_{\alg{B}}$
 iff $\rho (T(B))=\rho (B)$ for all $B\in \alg{B}$
and for all states $\rho$ of $\alg{A}\vee \alg{B}$. Since
 all states of $\alg{B}$ are
restrictions of states on $\alg{A}\vee \alg{B}$,
it follows that $(T^{*}\rho )|_{\alg{B}}=\rho |_{\alg{B}}$ if and only if
 $\omega (T(B))=\omega (B)$
for all states $\omega$ of $\alg{B}$, i.e., if and only if $T(B)=B$ for
all $B\in \alg{B}$.

It is clear that the kinematic independence of $\alg{A}$ and $\alg{B}$
entails that Alice's local measurement operations cannot convey
any
information to Bob (i.e., $T(B)= \\
\sum_{i=1}^{n}E_{i}^{1/2}BE_{i}^{1/2}= B$
 for $B\in \alg{B}$ if $T$ is implemented by a positive operator valued
resolution of the identity in \alg{A}).
Thus, we need only show that if Alice cannot convey any information
to Bob by performing local measurement operations, then $\alg{A}$
and $\alg{B}$ are kinematically independent.  In the standard Hilbert
space case, our argument would proceed as follows: Consider any ideal
(L{\"u}ders), non-selective measurement of the form
\begin{eqnarray} T(A) &=& PAP+(I-P)A(I-P) \qquad \qquad (A\in
  \alg{A}\vee \alg{B}) \label{luders} \end{eqnarray}
where $P$ is a projection in $\alg{A}$.  Then no superluminal
information transfer via measurement entails that
for any $B\in \alg{B}$, \begin{eqnarray}
B &=& T(B)\quad =\quad PBP+(I-P)B(I-P) \end{eqnarray}
and therefore \begin{eqnarray}
2PBP -PB-BP &=& 0 \end{eqnarray}  Multiplying on the left
(respectively, right) with $P$, and using the fact that
$P^{2}=P$, we then obtain $PBP-PB=0$ (respectively, $PBP-BP=0$).
Subtracting these two equations gives $[P,B]=0$.  Thus, since
$\alg{A}$ is spanned by its projections and
$\alg{B}$ is spanned by its self-adjoint operators, it follows
that $\alg{A}$ and $\alg{B}$ are kinematically independent.

In the more general $C^{*}$-algebraic framework, this argument is not
available: Since the algebra $\alg{A}$ does not necessarily
contain projection operators, we cannot assume that there are any
measurement operations of the form given in Eqn.~\ref{luders} where
$P$ is a projection. Instead, since C*-algebras are spanned by their
effects (positive operators),
consider the simplest
case of a POV measurement defined by
\begin{eqnarray} T_{E}(A) &=& E^{1/2}AE^{1/2}+(I-E)^{1/2}A(I-E)^{1/2}
   \qquad (A \in \alg{A}\vee \alg{B})  \end{eqnarray}
where $E$ is some effect in $\alg{A}$.  We will now show that
when $B$ is self-adjoint, $T_{E}(B)=B$ entails
that $[E,B]=0$.

\begin{thm} $T_{E}(B)=B$ for all effects $E \in \alg{A}$ and
  self-adjoint operators $B \in \alg{B}$ only if $\alg{A}$ and
  $\alg{B}$ are kinematically independent.
\end{thm}

For the proof of this theorem, recall that a derivation is a linear
map such that $d(AB)=A(dB)+(dA)B$.

\begin{proof}  Suppose that $T_{E}(B)=B$ where $E$ is an effect in
  $\alg{A}$ and $B$ is a self-adjoint operator in $\alg{B}$.  Then a
  tedious but elementary calculation shows that
\begin{eqnarray} [E^{1/2},[E^{1/2},B]] &=& 0  \end{eqnarray}
Clearly, the map $X\mapsto i[E^{1/2},X]$ defines a derivation $d$ of
$\alg{A}\vee \alg{B}$.  Moreover, since $d(dB)=0$ and $B$ is
self-adjoint, it follows that $i[E^{1/2},B]=dB=0$ \cite[Appendix
A]{Busch}.  Thus, $[E,B]=0$.  Finally, since a $C^{*}$-algebra is
spanned by its effects, if $T_{E}(B)=B$ for all effects $E\in \alg{A}$
and self-adjoint operators $B\in \alg{B}$, then $\alg{A}$ and
$\alg{B}$ are kinematically independent.  \end{proof}

Thus, the kinematic independence of $\alg{A}$ and $\alg{B}$ is
equivalent to the `no superluminal information transfer by
measurement' constraint. In deriving our
subsequent results, we assume kinematic independence.

\subsection{No Broadcasting and Noncommutativity}

In a cloning process, a ready state $\sigma$ of system B, and the
state to be cloned $\rho$ of system A, are transformed into two copies
of $\rho$.  Thus, such a process creates no correlations between the
states of A and B.  By contrast, in a more general broadcasting
process, a ready state $\sigma$, and the state to be broadcast $\rho$
are transformed to a new state $\omega$ of A+B, where the marginal
states of $\omega$ with respect to both A and B are
$\rho$.\footnote{We are indebted to Rob Spekkens for clarifying this
  distinction for us, and for supplying relevant references.}  In the
context of elementary quantum mechanics, neither cloning nor
broadcasting is generally possible: A pair of pure states can be
cloned if and only if they are orthogonal (Wootters and Zurek
 \cite{Wootters}, Dieks \cite{Dieks}), and
(more generally) an arbitrary pair of states can be broadcast if and
only if they are represented by mutually commuting density matrices
(Barnum \textit{et al} \cite{Barnum}).
Thus, one might suspect that in a classical theory (in
which all operators commute), all states can be broadcast.  In this
section, we show that this is indeed the case; and, in fact, the
ability to broadcast states distinguishes classical systems from
quantum systems.

We now introduce a general notion of broadcasting for a pair
$\alg{A},\alg{B}$ of kinematically independent $C^{*}$-algebras.  But
we must first establish the existence and uniqueness of product states
of $\alg{A}\vee \alg{B}$.

A state $\rho$ of $\alg{A}\vee \alg{B}$ is said to be a \emph{product
  state} just in case $\rho (AB)=\rho (A)\rho (B)=\rho (BA)$ for all
$A\in \alg{A}$ and $B\in \alg{B}$.  Since $\alg{A}$ and $\alg{B}$ are
both $C^{*}$-independent and kinematically independent, there is an
isomorphism $\pi$ from the $*$-algebra generated by $\alg{A}$ and
$\alg{B}$ onto the algebraic tensor product $\alg{A}\odot \alg{B}$
such that $\pi (AB)=A\otimes B$ for all $A\in \alg{A}$ and $B\in
\alg{B}$.  We will occasionally omit reference to $\pi$ and use
$A\otimes B$ to denote the product of $A\in \alg{A}$ and $B\in
\alg{B}$.  It also follows that $\pi$ can be extended to a continuous
surjection $\overline{\pi}$ from $\alg{A}\vee \alg{B}$ onto the
spatial tensor product $\alg{A}\otimes \alg{B}$ \cite[Thm.~1]{Florig}.
Since the state space of $\alg{A}\otimes \alg{B}$ has a natural tensor
product structure, we can use the mapping $\overline{\pi}$ to define
product states of $\alg{A}\vee \alg{B}$.  In particular, for each
state $\rho$ of $\alg{A}\otimes \alg{B}$ define the state $\pi
^{*}\rho$ of $\alg{A}\vee \alg{B}$ by setting
\begin{eqnarray}
(\pi ^{*}\rho )(A) &=& \rho (\pi (A)) \qquad (A \in \alg{A}\vee
\alg{B}) \end{eqnarray}
If $\omega$ is a state of $\alg{A}$ and $\rho$ is a state of
$\alg{B}$, then $\pi
^{*}(\omega \otimes \rho)$ is a product state of $\alg{A}\vee
\alg{B}$ with marginal states $\omega$ and $\rho$.  In fact, the
following result shows that $\pi ^{*}(\omega \otimes \rho)$ is the
unique product state of $\alg{A}\vee \alg{B}$ with these marginal
states.

\begin{lemma} Suppose that $\alg{A}$ and $\alg{B}$ are kinematically
  independent $C^{*}$-algebras.  Then for any state $\omega$ of
  $\alg{A}$ and for any state $\rho$ of $\alg{B}$ there is at most one
  state $\sigma$ of $\alg{A}\vee \alg{B}$ such that $\sigma
  (AB)=\omega (A)\rho (B)$ for all $A\in \alg{A}$ and $B\in \alg{B}$.
  \label{unique}
\end{lemma}

\begin{proof} The set $\alg{S}$ of finite sums of the form $\sum
  _{i=1}^{n}A_{i}B_{i}$ with $A_{i}\in \alg{A}$ and $B_{i}\in \alg{B}$
  is a $^{*}$-algebra containing both $\alg{A}$ and $\alg{B}$.  Moreover,
  $\alg{S}$ is clearly contained in any $^{*}$-algebra that contains both
  $\alg{A}$ and $\alg{B}$.  Thus, $\alg{S}$ is the $^{*}$-algebra
  generated by $\alg{A}$ and $\alg{B}$.  Suppose then that $\sigma _{0}$
  and $\sigma _{1}$ are states of $\alg{A}\vee \alg{B}$ such that
  $\sigma _{0}(AB)=\omega (A)\rho (B)=\sigma _{1}(AB)$ for all $A\in
  \alg{A}$ and $B\in \alg{B}$.  Then \begin{eqnarray} \sigma
    _{0}\left( \sum _{i=1}^{n}A_{i}B_{i}\right) &=& \sum
    _{i=1}^{n}\omega (A_{i})\rho (B_{i}) \quad =\quad \sigma
    _{1}\left( \sum _{i=1}^{n}A_{i}B_{i}\right)  \end{eqnarray} for
  all $A_{i}\in \alg{A}$ and $B_{i}\in \alg{B}$.  Since $\alg{A}\vee
  \alg{B}$ is the closure of $\alg{S}$ in the norm topology, and since
  states are continuous in the norm topology, $\sigma _{0}=\sigma
  _{1}$.  \end{proof}

When it will not cause confusion, we will henceforth supress reference
to the state mapping $\pi ^{*}$, so that $\omega \otimes \rho$ denotes
the unique product state on $\alg{A}\vee \alg{B}$ with marginals
$\omega$ and $\rho$.

\begin{defn} Given two isomorphic, kinematically independent
  $C^{*}$-algebras $\alg{A}$ and $\alg{B}$, we say that a pair $\{\rho
  _{0}, \rho _{1}\}$ of states of $\alg{A}$ can be \emph{broadcast}
  just in case there is a standard state $\sigma$ of $\alg{B}$ and a
  dynamical evolution represented by an operation $T$ of $\alg{A}\vee
  \alg{B}$ such that $T^{*}(\rho _{i} \otimes \sigma )|_{\alg{A}} =
  T^{*}(\rho _{i}\otimes \sigma )|_{\alg{B}} = \rho _{i} \; (i =
  0,1)$. We say that a pair $\{\rho _{0}, \rho_{1}\}$ of states of
  $\alg{A}$ can be \emph{cloned} just in case $T^{*}(\rho _{i} \otimes
  \sigma ) = \rho _{i} \otimes \rho _{i} \; (i = 0,1)$.
\end{defn}

We show next that pairwise broadcasting is always possible in
classical systems.  Indeed, when the algebras of observables are
abelian, there is a `universal' broadcasting map that clones any pair
of input pure states and broadcasts any pair of input mixed states.

\begin{thm} If $\alg{A}$ and $\alg{B}$ are abelian then there is an
  operation $T$ on $\alg{A}\vee \alg{B}$ that broadcasts all states of
  $\alg{A}$. \end{thm}

\begin{proof} Since $\alg{A}$ is abelian, $\alg{A}\vee \alg{B}$ is
  naturally isomorphic to $\alg{A}\otimes \alg{B}$
  \cite[Theorem~2]{Roos}.  Since $\alg{A}$ and $\alg{B}$ are
  isomorphic abelian algebras, both are isomorphic to the space $C(X)$,
  where $X$ is some compact Hausdorff
  space, and therefore $\alg{A}\otimes \alg{B}\cong C(X)\otimes
  C(X)\cong C(X\times X)$ \cite[p.~849]{Kadison}.  Define a mapping
  $\eta$ from $X\times X$ into $X\times X$ by setting
\begin{equation}
\eta ((x,y))=(x,x) \qquad \qquad (x,y\in X)\end{equation}  Since
$\eta$ is continuous, we can define a linear mapping $T$ on
$C(X\times X)$ by
setting $Tf=f\circ \eta$.  Since the range of $Tf$ is a subset of the
range of $f$, the mapping $T$ is positive, and $T(I)=I$.  Furthermore,
every positive mapping whose domain or range is an abelian algebra is
completely
positive \cite[Exercise 11.5.22]{Kadison}.  Therefore, $T$ is a
nonselective operation on $C(X)\otimes C(X)$.  To see that $T$
broadcasts all states, note first that
$T(f\otimes g)=fg\otimes I$ for any product function $f\otimes g$.  In
particular, $T(I\otimes f)=f\otimes I$, and thus
\begin{eqnarray} T^{*}(\rho \otimes \sigma )(I\otimes f) &=& (\rho
  \otimes \sigma )(f\otimes I) \quad = \quad \rho (f) \end{eqnarray}
for any states $\rho ,\sigma$ of $C(X)$.  That is,
$T^{*}(\rho \otimes \sigma )|_{\alg{B}}=\rho$.  On the other
hand, since $T(f\otimes I)=f\otimes I$, it follows that
\begin{eqnarray} T^{*}(\rho \otimes \sigma )(f\otimes I) &=& (\rho
  \otimes \sigma )(f\otimes I) \quad =\quad \rho (f)
  \end{eqnarray} for any states $\rho ,\sigma$ of $C(X)$.  That is,
$T^{*}(\rho \otimes \sigma )|_{\alg{A}}=\rho$.
\end{proof}

Barnum \textit{et al} \cite{Barnum} note that if density operators
$D_{0},D_{1}$ on a Hilbert space $\hil{H}$ can be simultaneously
diagonalized, then there is a unitary operator $U$ that broadcasts the
corresponding pair of states.  Thus, such states can be broadcast by a
\emph{reversible} operation, and the added strength of
\emph{irreversible} (general completely positive) operations is not
necessary.  However, the broadcasting operation $T$ defined in the
previous theorem is patently irreversible, since it corresponds to a
many-to-one mapping $(x,y)\mapsto (x,x)$ of the pure state space.
Indeed, although there are many physically significant classical
systems where broadcasting can be performed via reversible operations,
this is not generally true.  Consider the following two contrasting
cases.

First, in the case of classical particle mechanics, systems A and B
each have the phase space $\mathbb{R}^{6n}$, for some finite $n$.  Thus,
$\alg{A}\vee \alg{B}\cong C(\mathbb{R}^{6n})\otimes
C(\mathbb{R}^{6n})\cong C(\mathbb{R}^{6n}\times \mathbb{R}^{6n})$,
where $C(\mathbb{R}^{6n})$ is the set of bounded continuous functions
from $\mathbb{R}^{6n}$ into $\mathbb{C}$.  Let the ready state of
system B be the zero vector in $\mathbb{R}^{6n}$, and let $\eta$ be
the invertible linear transformation of $\mathbb{R}^{6n}\times
\mathbb{R}^{6n}$ given by the matrix
\[ \left( \begin{array}{cc} I & -I \\
I & I \end{array} \right)  \]
where $I$ is the identity matrix of $\mathbb{R}^{6n}$.  Since $\eta$
is an autohomeomorphism of $\mathbb{R}^{6n}\times \mathbb{R}^{6n}$,
the mapping $f\mapsto f\circ \eta$ defines an automorphism of
$\alg{A}\vee \alg{B}$ \cite[Thm.~3.4.3]{Kadison}.  Moreover,
$
\eta ((\mathbf{x},\mathbf{0})) = (\mathbf{x},\mathbf{x})
$
for any pure state $\mathbf{x}$, and an argument similar to that used
above shows that $\eta$ broadcasts arbitrary states.

Second, suppose that systems A and B each have the phase space
$\mathbb{N}^{*}=\mathbb{N}\cup \{ \infty \}$, where the open sets of
$\mathbb{N}^{*}$ consist of all finite subsets of $\mathbb{N}$, plus
all cofinite sets containing $\infty$.  (That is, $\mathbb{N}^{*}$ is
the one-point compactification of $\mathbb{N}$.)  Then $\alg{A}\vee
\alg{B}\cong C(\mathbb{N}^{*})\otimes C(\mathbb{N}^{*})\cong C(\onept
)$, and every automorphism $\alpha$ of $\alg{A}\vee \alg{B}$ is
induced by an autohomeomorphism $\eta$ of $\onept$ via the equation
$\alpha (f)=f\circ \eta$ \cite[Thm.~3.4.3]{Kadison}.  However, it is
not difficult to see that there is no autohomeomorphism of $\onept$
that clones arbitary pairs of pure states.  In particular, for any
$n\in \mathbb{N}^{*}$, there is an $m\in \mathbb{N}^{*}$ such that
$(n,m)$ is not mapped onto $(m,m)$ by any autohomeomorphism of
$\onept$.  Therefore, general classical systems do not permit
broadcasting via a reversible operation.

We now show that general quantum systems do not permit
broadcasting.  In particular, we prove that if any two states of a
system can be broadcast, then that system has an abelian algebra of
observables.  Our proof proceeds by showing
that if any two states can be broadcast, then any two
\emph{pure} states can be cloned; and that if two pure states of a
$C^{*}$-algebra can be cloned, then they must be orthogonal.

Two pure states $\rho ,\omega$ of a $C^{*}$-algebra are said to be
orthogonal just in case $\norm{\rho -\omega }=2$.  More generally, the
transition probability $p(\rho ,\omega )$ is defined to be (see
\cite{Roberts}):
\begin{eqnarray} p(\rho ,\omega ) &=& \textstyle 1-\frac{1}{4}\norm{
    \rho -\omega }^{2} \end{eqnarray}
If $\rho$ is a state of $\alg{A}$, and $U$ is a unitary operator in
$\alg{A}$, then we let $\rho _{U}$ denote the state defined by
\begin{eqnarray}
\rho _{U}(A) &=& \rho (U^{*}AU) \qquad \qquad (A\in \alg{A})
\end{eqnarray}
In this case, $\rho$ and $\rho _{U}$ are said to be unitarily
  equivalent.  Furthermore, if $U$ is a unitary operator in $\alg{A}$
and $V$ is a unitary operator in $\alg{B}$, then
\begin{eqnarray} (\omega \otimes \rho )_{U\otimes V}(A\otimes B) &=&
  (\omega \otimes \rho )(U^{*}AU\otimes V^{*}BV) \\
&=& (\omega _{U}\otimes \rho _{V})(A\otimes B) \end{eqnarray}
for all $A\in \alg{A}$ and $B\in \alg{B}$.  Thus, the uniqueness of
product states (Lemma~\ref{unique}) entails that $(\omega \otimes
  \rho )_{U\otimes V}=\omega _{U}\otimes \rho _{V}$.

  For the following lemma, we will need to make use of the fact that
  $p(\rho ,\rho _{U})=\abs{\rho (U)}^{2}$ for any pure state $\rho$
  \cite[Lemma 2.4]{Powers}.

\begin{lemma} If $\rho _{0},\rho _{1}$ are unitarily equivalent pure
  states of $\alg{A}$, and $\sigma$ is an arbitrary state of
  $\alg{B}$ then:
\begin{eqnarray}  p(\rho _{0}\otimes \sigma ,\rho _{1}\otimes \sigma
  ) &=& p(\rho _{0},\rho _{1}) \\
p  (\rho _{0}\otimes \rho _{0},\rho _{1}\otimes \rho
  _{1}) &=& p(\rho _{0},\rho _{1})^{2}  \end{eqnarray}
\label{duplicate}
  \end{lemma}

\begin{proof} Since $\rho _{0}$ and $\rho _{1}$ are unitarily
equivalent, there is a
  unitary operator $U\in \alg{A}$ such that $\rho _{1}=(\rho
  _{0})_{U}$ and $p(\rho _{0},\rho _{1})=\abs{\rho _{0}(U)}^{2}$.
  Thus $\rho _{1}\otimes \sigma =(\rho _{0}\otimes \sigma )_{(U\otimes
    I)}$, and therefore
\begin{equation} p(\rho _{0}\otimes \sigma ,\rho _{1}\otimes \sigma
  )\:=\:\abs{(\rho _{0}\otimes \sigma )(U\otimes I)}^{2}\:=\:\abs{\rho
    _{0}(U)}^{2}
\:=\:p(\rho _{0},\rho _{1}) \end{equation}
Similarly, $\rho _{0}\otimes \rho _{0}= (\rho
_{1}\otimes \rho _{1})_{(U\otimes U)}$, and therefore
\begin{equation}
p( \rho _{0}\otimes \rho _{0},\rho _{1}\otimes \rho _{1}) \:=\: \abs{
  (\rho _{0}\otimes \rho _{0})(U\otimes U)}^{2} \:=\: \abs{ \rho
_{0}(U) }^{4} \:=\:
p(\rho _{0},\rho _{1})^{2}  \end{equation} \end{proof}

\begin{lemma} Suppose that $\alg{A}$ and $\alg{B}$ are kinematically
  independent.  If $\rho$ is a state of $\alg{A}\vee \alg{B}$ such
  that $\rho |_{\alg{A}}$ is pure or $\rho |_{\alg{B}}$ is pure, then
  $\rho$ is a product state.
  \label{pure-reduction} \end{lemma}

\begin{proof} Let $\omega =\rho |_{\alg{A}}$, and let $B$ be an effect
  in $\alg{B}$.  Define positive linear functionals $\lambda _{1}$ and
  $\lambda _{2}$ on $\alg{A}$ by setting
\begin{eqnarray}
\lambda _{1}(A) &=& \rho (B^{1/2}AB^{1/2}) \quad =\quad \rho (AB) \\
\lambda _{2}(A) &=& \rho ((I-B)^{1/2}A(I-B)^{1/2}) \quad =\quad \rho
(A(I-B)) \end{eqnarray}
for all $A\in \alg{A}$.  It then follows that \begin{eqnarray}
\omega (A) &=& \rho (A) \quad =\quad \lambda _{1}(A)+\lambda _{2}(A)
\quad \geq \quad \lambda _{1}(A) \end{eqnarray}  Since $\omega$ is a
pure state of
$\alg{A}$, $\lambda _{1}$ is a nonnegative multiple $k\omega$ of
$\omega$, and
\begin{eqnarray} k &=& k\omega (I)\quad =\quad \lambda _{1}(I)\quad
  =\quad \rho (B) \end{eqnarray} Accordingly,
\begin{eqnarray}
\rho (AB) &=& \lambda _{1}(A) \quad =\quad k\omega (A) \quad =\quad
\omega (A)\rho (B)
\end{eqnarray} By linearity, the same equation holds when we replace
$B$ by an arbitrary element of $\alg{B}$.  Therefore, $\rho
(AB)=\omega (A)\rho
(B)=\rho (A)\rho (B)$ for all $A\in \alg{A}$ and $B\in \alg{B}$.
\end{proof}

The proof of the next theorem turns on the fact that nonselective
operations cannot \emph{increase} the norm distance between states,
and therefore cannot \emph{decrease} the transition probabilities
between states.  That is, for any nonselective operation $T$,
\begin{eqnarray} p(T^{*} \omega ,T^{*}\rho ) &\geq& p(\omega ,\rho )
  \end{eqnarray}
for all states $\omega ,\rho$.  To see this, note that
\begin{eqnarray}
\norm{ T^{*}\omega - T^{*}\rho } &=& \sup \bigl\{ \, \abs{ (\omega -
\rho
  )(T(A))} \; : \; \norm{A}\leq 1 \, \bigr\} \end{eqnarray}
and the Russo-Dye theorem entails that if $\norm{A}\leq 1$ then
$\norm{T(A)}\leq
  1$.

\begin{thm} If for each pair $\{ \rho _{0},\rho _{1}\}$ of states
  of $\alg{A}$, there is an operation $T$ on $\alg{A}\vee \alg{B}$
  that broadcasts $\{ \rho _{0},\rho _{1} \}$, then $\alg{A}$ is
  abelian.  \end{thm}

\begin{proof} We assume that for each pair $\{ \rho _{0},\rho _{1}\}$
of
  states of $\alg{A}$, there is an operation $T$ on $\alg{A}\vee
  \alg{B}$ that broadcasts $\{ \rho _{0},\rho _{1}\}$.  Suppose for
  reductio ad absurdum that $\alg{A}$ is not abelian.  Then there are
  pure states $\rho _{0},\rho _{1}$ of $\alg{A}$ such that
  $0<\norm{\rho _{0}-\rho _{1}}<2$ \cite[Exercise 4.6.26]{Kadison}.
  In this case, $\rho _{0}$ and $\rho _{1}$ are unitarily equivalent
  \cite[Corollary 10.3.8]{Kadison}.  By hypothesis, there is a
  standard state $\sigma$ of $\alg{B}$ and an operation $T$ on
  $\alg{A}\vee \alg{B}$ such that
\begin{eqnarray}
T^{*}(\rho _{0}\otimes \sigma )|_{\alg{A}} &=& T^{*}(\rho _{0}\otimes
\sigma )|_{\alg{B}} \quad = \quad \rho _{0} \\
T^{*}(\rho _{1}\otimes \sigma )|_{\alg{A}} &=& T^{*}(\rho _{1}\otimes
\sigma )|_{\alg{B}} \quad = \quad \rho _{1}  \end{eqnarray}
Since $\rho _{0}$ and $\rho _{1}$ are pure, it follows from
Lemma~\ref{pure-reduction} that $T^{*}(\rho _{0}\otimes
\sigma )=\rho _{0}\otimes \rho _{0}$ and $T^{*}(\rho _{1}\otimes
\sigma )=\rho _{1}\otimes \rho _{1}$.
Thus, \begin{eqnarray} p(\rho _{0},\rho _{1}) &=& p(\rho _{0}\otimes
  \sigma ,\rho _{1}\otimes \sigma ) \label{first} \\
  &\leq & p(T^{*}(\rho _{0}\otimes \sigma ) ,T^{*}(\rho _{1}\otimes
  \sigma )) \label{decrease} \\
  &=& p(\rho _{0}\otimes \rho _{0},\rho _{1}\otimes \rho _{1} ) \\
  &=& p(\rho _{0},\rho _{1})^{2} \label{second} \end{eqnarray} (The
equalities in
Equations~\ref{first} and~\ref{second} follow from
Lemma~\ref{duplicate}, while the inequality in Equation~\ref{decrease}
follows from the fact that transition probabilities cannot decrease
under $T^{*}$.)  However, the inequality $p(\rho _{0},\rho _{1})\leq
p(\rho _{0},\rho _{1})^{2}$ contradicts the fact that $0<p(\rho
_{0},\rho _{1})<1$.
Therefore, $\alg{A}$ is abelian.  \end{proof}

\subsection{No Bit Commitment and Nonlocality}

We show that the impossibility of unconditionally secure bit
commitment between systems A and B, in the presence of the kinematic
independence and noncommutativity of their algebras of observables,
entails nonlocality: spacelike separated systems must at least
sometimes occupy entangled states.  Specifically, we show that if
Alice and Bob have spacelike separated quantum systems, but cannot
prepare any entangled state, then Alice and Bob can devise an
unconditionally secure bit commitment protocol.

We first show that quantum systems are characterized by the existence
of non-uniquely decomposable mixed states.

\begin{lemma} Let $\alg{A}$ be a $C^{*}$-algebra.  Then $\alg{A}$ is
  nonabelian if and only if there are distinct pure states
  $\omega_{1,2}$ and $\omega_{\pm}$ of $\alg{A}$ such that
  $(1/2)(\omega_{1}+\omega_{2})=(1/2)(\omega_{+}+\omega_{-})$.
\label{mixtures} \end{lemma}

\begin{proof} If $\alg{A}$ is abelian then its states are
  in one-to-one correspondence with measures on its pure state space.
  In particular, if $\rho =(1/2)(\omega _{1}+\omega _{2})$, where
  $\omega _{1},\omega _{2}$ are distinct pure states, then this
  decomposition is unique.

  Conversely, suppose that there are $A,B\in \alg{A}$ such that
  $[A,B]\neq 0$ (cf.~\cite[Example 4.2.6]{Bratteli}) .  Then there is
  a pure state $\rho$ of $\alg{A}$ such that $\rho([A,B])\neq 0$
  \cite[Thm.~4.3.8]{Kadison}. Let $(\pi ,\hil{H},\Omega )$ be the GNS
  representation of $\alg{A}$ induced by $\rho$.  The dimension of the
  Hilbert space $\hil{H}$ must exceed one; for otherwise
\begin{equation} \pi ([A,B])=[\pi (A),\pi (B)]=0  \end{equation} in
contradiction with the fact that $\langle \Omega ,\pi ([A,B])\Omega
\rangle= \rho([A,B])\neq 0$.  Thus, there is a pair $x_{1},x_{2}$ of
orthogonal unit vectors in $\hil{H}$.  Define two states $\omega_{i}$
of $\alg{A}$ by setting
\begin{eqnarray}
\omega_{i}(A)&=& \langle x_{i},\pi (A)x _{i} \rangle \qquad \qquad
(A\in\alg{A}) \end{eqnarray}
Similarly, define, in the same way, another pair of states
$\omega_{\pm}$ of $\alg{A}$ using the orthogonal unit
vectors $2^{-1/2}(x_{1}\pm x_{2})$.  Since $\rho$ is pure,
$(\pi ,\hil{H},\Omega )$ is irreducible, and all four of the states
$\omega_{1,2}$ and $\omega_{\pm}$ are pure and distinct.  Moreover, by
construction
$(1/2)(\omega_{1}+\omega_{2})=(1/2)(\omega_{+}+\omega_{-})$.
\end{proof}

If each of A and B has a non-uniquely decomposable mixed state, then
A+B has a pair $\{ \rho _{0}, \rho _{1}\}$ of distinct (classically)
correlated states whose marginals relative to A and B are identical.
In particular, let
\begin{eqnarray} \rho _{0}&=& (1/2)\left( \omega _{1}\otimes
    \omega
    _{1}+\omega _{2}\otimes \omega _{2}\right) \label{zero} \\
  \rho _{1}&=&  (1/2)\left( \omega _{+}\otimes \omega _{+}+\omega
    _{-}\otimes \omega _{-}\right) \label{one} \end{eqnarray}
The protocol then proceeds as follows: Alice and Bob arrange things so
    that at the commitment stage, Alice can make a choice that will
    determine that either $\rho _{0}$ or $\rho _{1}$ is prepared as a
    suitably long sequence of pure states ($\omega _{1},\omega _{2}$
    or $\omega _{+},\omega _{-}$), the
    former corresponding to the commitment~$0$, and
    the latter corresponding to the commitment~$1$.  Alice and Bob
    also agree that at the revelation stage, if Alice committed to $0$ then
    she will instruct Bob to perform a measurement that will
    distinguish between states $\omega _{1}$ and $\omega _{2}$, and if
    she committed to $1$ then she will instruct Bob to perform a
    measurement that will distinguish between states $\omega _{+}$ and
    $\omega _{-}$.  We must be cautious on this last point:
    neither Alice nor Bob will typically be able to perform a
    measurement that can discriminate with certainty between these
    states.  However, for any $\epsilon >0$, there is an effect
    $A\in \alg{A}$ such that $\omega _{1}(A)>1-\epsilon$ and $\omega
    _{2}(A)<\epsilon$.  Similarly, there is an effect $B\in \alg{A}$
    such that $\omega _{+}(B)>1-\epsilon$ and $\omega
    _{-}(B)<\epsilon$.  That is, Alice and Bob
    can perform measurements that will discriminate with arbitrary
    accuracy between $\omega _{1}$ and $\omega _{2}$, or between
    $\omega _{+}$ and $\omega _{-}$.  Finally, Alice will verify her
    commitment by performing the corresponding measurement on
    her system and reporting the outcomes to Bob.

    Our final theorem shows that if Alice and Bob have access only to
classically correlated states (i.e., convex combinations of product
states), then this bit commitment protocol is secure.  In particular,
we show that Alice cannot cheat by preparing some state $\sigma$ which
she could later transform at will into either $\rho _{0}$ or $\rho
_{1}$.  To be precise, the no superluminal information transfer by
measurement constraint entails that Alice can perform an operation $T$
on $\alg{A}\vee \alg{B}$ only if $T(B)=B$ for all $B\in \alg{B}$, and
$T(A)\in \alg{A}$ for all $A\in \alg{A}$.  It follows then that Alice
can transform product states only to other product states.

\begin{thm} If $\alg{A}$ and $\alg{B}$ are nonabelian then there is a
  pair $\{ \rho _{0},\rho _{1}\}$ of states of $\alg{A}\vee \alg{B}$
  such that:
\begin{enumerate}
\item $\rho _{0}|_{\alg{B}}=\rho _{1}|_{\alg{B}}$.
\item There is no classically correlated state $\sigma$ of
  $\alg{A}\vee \alg{B}$ and operations $T_{0}$ and $T_{1}$ performable
  by Alice such that $T_{0}^{*}\sigma=\rho _{0}$ and
  $T^{*}_{1}\sigma=\rho _{1}$.
\end{enumerate}
\end{thm}

For the proof of this theorem, we recall that two representations
$(\pi ,\hil{H})$ and $(\phi ,\hil{K})$ of a $C^{*}$-algebra are said
to be quasi-equivalent just in case there is a $*$ isomorphism
$\alpha$ from $\pi (\alg{A})''$ onto $\phi (\alg{A})^{''}$ such that
$\alpha (\pi (A))=\phi (A)$ for each $A$ in $\alg{A}$.  Similarly,
states $\omega$ and $\rho$ of $\alg{A}$ are said to be
quasi-equivalent just in case their corresponding GNS representations
are quasi-equivalent.  Finally, quasi-equivalence is an equivalence
relation, and is closed under finite convex combinations.

\begin{proof} Let $\rho _{0}$ and $\rho _{1}$ be the states defined in
  Eqns.~\ref{zero} and \ref{one}.  Suppose that $\sigma =\sum
  _{i=1}^{n}\lambda _{i}(\alpha _{i}\otimes \beta _{i})$, where the
  $\alpha _{i}$ are states of $\alg{A}$, and the $\beta _{i}$ are
  states of $\alg{B}$.  Let $T_{0}$ and $T_{1}$ be operations of
  $\alg{A}\vee \alg{B}$ that can be performed by Alice.  Then for each
  $i\in [1,n]$, there are states $\alpha '_{i}$ and $\alpha ''_{i}$ of
  $\alg{A}$ such that $T^{*}_{0}(\alpha _{i}\otimes \beta _{i})=\alpha
  _{i}'\otimes \beta _{i}$ and $T^{*}_{1}(\alpha _{i}\otimes \beta
  _{i})=\alpha ''_{i}\otimes \beta _{i}$.  Moreover, since $T^{*}_{0}$
  and $T^{*}_{1}$ are affine,
\begin{eqnarray}
\rho _{0} &=& T^{*}_{0}\sigma \quad =\quad \sum _{i=1}^{n}\lambda
_{i}(\alpha _{i}'\otimes
\beta _{i}) \label{manchester} \\
\rho _{1} &=& T^{*}_{1}\sigma \quad =\quad \sum _{i=1}^{n}\lambda
_{i}(\alpha _{i}''\otimes \beta _{i}) \label{united} \end{eqnarray}
Let $\mu$ denote the mixed state $(1/2)(\omega _{1}+\omega
_{2})=(1/2)(\omega _{+}+\omega
_{-})$ of $\alg{B}$.  Then $\sum
_{i=1}^{n}\lambda _{i}\beta _{i}=\mu$, so that each $\beta _{i}$ is
quasi-equivalent to $\mu$.  Let $(\pi ,\hil{H})$
be the representation of $\alg{B}$ defined in
Lemma~\ref{mixtures}, let $P_{1}$ denote the projection onto $x_{1}$,
and let $P_{+}$ denote the projection onto $2^{-1/2}(x_{1}+x_{2})$.
(Note that since $\omega _{1,2},\omega _{\pm}$ are
represented by vectors in $\hil{H}$, it follows that $(\pi ,\hil{H})$
is unitarily equivalent to the GNS representations induced by these
states.  Moreover, $(\pi ,\hil{H})$ is quasi-equivalent to the GNS
representation induced by $\beta _{i}$.)  Since $\pi
(\alg{B})$ is weakly dense in $\mathbf{B}(\hil{H})$, there are nets
$\{ A_{i}\}\subseteq \alg{B}$ and $\{ B_{i}\}\subseteq \alg{B}$ such
that $\pi (A_{i})$ converges ultraweakly to $P_{1}$ and $\pi (B_{i})$
converges ultraweakly to $P_{+}$.  (Moreover, we can choose these nets
so that $0\leq \pi (A_{i}),\pi (B_{i})\leq I$ for all $i$.)  Since
each of the states $\omega _{1,2},\omega _{\pm}$
is represented by a vector in $\hil{H}$, ultraweak continuity of normal
states entails that:
\begin{eqnarray}
\lim _{i} \rho _{1}(A_{i}\otimes (I-A_{i})) &=& 0  \\
\lim _{i} \rho _{1}((I-A_{i})\otimes A_{i}) &=& 0  \\
\lim _{i}\rho _{0}(B_{i}\otimes (I-B_{i})) &=& 0  \\
\lim _{i}\rho _{0}((I-B_{i})\otimes B_{i}) &=& 0  \end{eqnarray}
Furthermore, since $\lim _{i}\mu (A_{i})=1/2$, there exists some $j\in
[1,n]$ such that \mbox{$\lim _{i}\beta _{j}(A_{i})>0$}.   Let $\beta
=\beta
_{j}$ and let $\alpha '=\alpha
_{j}'$.  Then, combining the previous equalities with
Eqns.~\ref{manchester} and
\ref{united} gives:
\begin{eqnarray}
\lim _{i}\alpha '(A_{i})\beta (I-A_{i}) &=& 0  \\
\lim _{i}\alpha '(I-A_{i})\beta (A_{i}) &=& 0  \\
\lim _{i}\alpha ''(B_{i})\beta (I-B_{i}) &=& 0  \\
\lim _{i}\alpha ''(I-B_{i})\beta (B_{i}) &=& 0 \end{eqnarray}
Since $0\leq \beta (A_{i}),\alpha '(A_{i})\leq 1$ for all $i$,
it follows that $\{ \beta (A_{i}) \}$ and $\{ \alpha '(A_{i})\}$ have
accumulation
points.  Thus, we may
pass to a subnet in which $\lim _{i}\beta (A_{i})$ and $\lim
_{i}\alpha '(A_{i})$ exist (and the preceding equations still hold).
We now claim that $\lim _{i}\beta (A_{i})=1$.  Indeed, since
$\lim _{i}\alpha (I-A_{i})\beta (A_{i})=0$, if $\lim _{i}\beta
(A_{i})>0$ then $1-\lim _{i}\alpha '(A_{i})=\lim _{i}\alpha
'(I-A_{i})=0$.
Moreover, since $\lim \alpha '(A_{i})\beta (I-A_{i})=0$, it follows
that that $1-\lim _{i}\beta (A_{i})=\lim _{i}\beta (I-A_{i})=0$.  Thus
$\lim _{i}\beta (A_{i})=1$.  An analogous argument
shows that either $\lim _{i}\beta (B_{i})=0$ or $\lim _{i}\beta
(B_{i})=1$.

Now, since the GNS representation induced by $\beta$ is
quasi-equivalent to the irreducible representation $(\pi ,\hil{H})$,
there is a density operator $D$ on $\hil{H}$ such that $\beta
(X)=\mathrm{Tr}(DX)$ for all $X\in \alg{B}$.  Since density operator
states are ultraweakly continuous, \begin{eqnarray}
  \mathrm{Tr}(DP_{1}) &=& \lim _{i}\mathrm{Tr}(DA_{i}) \quad =\quad 1
  ,\end{eqnarray} and therefore $D=P_{1}$.  Thus, if $\lim _{i}\beta
(B_{i})=0$ then we have a contradiction: \begin{eqnarray} 1/2 &=&
  \mathrm{Tr}(P_{1}P_{+}) \quad =\quad \mathrm{Tr}(DP_{+}) \quad
  =\quad \lim _{i}\mathrm{Tr}(DB_{i}) \quad =\quad 0.\end{eqnarray}
But $\lim _{i}\beta (B_{i})=1$ would also result in the contradiction
$1/2=1$.  Therefore, there is no classically correlated state $\sigma$
such that $T_{0}^{*}\sigma =\rho _{0}$ and $T^{*}_{1}\sigma =\rho
_{1}$.  \end{proof}

It follows that the impossibility of unconditionally secure bit
commitment entails that if each of a pair of separated physical
systems A and B has a non-uniquely decomposable mixed state, so that
A+B has a pair $\{ \rho _{0}, \rho _{1}\}$ of distinct classically
correlated states whose marginals relative to A and B are identical
(as in (\ref{zero}) and (\ref{one})), then A and B must be able to
occupy an entangled state that can be transformed to $\rho _{0}$ or
$\rho _{1}$ at will by a local operation.  The converse result remains
open: it is not known whether nonlocality---the fact that spacelike
separated systems occupy entangled states---entails the impossibility
of unconditionally secure bit commitment.  As we indicated in the
introduction, the proof of the corresponding result in elementary
quantum mechanics (in which all algebras are type I von Neumann
factors) depends on the biorthogonal decomposition theorem, via the
theorem of Hughston, Jozsa, and Wootters \cite{Hughston}.  Thus,
proving the converse would amount to generalizing the
Hughston-Jozsa-Wootters result to arbitrary nonabelian
$C^{*}$-algebras.  If, as we believe, the more general result holds,
then quantum theory can be characterized in terms of our three
information-theoretic constraints.  So, either quantum theory can be
characterized in terms of our information-theoretic constraints, or
there are physical systems which permit an unconditionally secure bit
commitment protocol.

\section{Concluding Remarks}

Within the framework of a class of theories broad enough to include
both classical and quantum particle and field theories, and hybrids
of
these theories, we have shown that
three information-theoretic constraints suffice to exclude the
classical theories. Specifically, the information-theoretic
constraints entail that the algebras of observables of distinct
physical systems commute, that the algebra of observables of each
individual system is noncommutative, and that spacelike separated
systems occupy entangled states.

Conversely, from the three physical characteristics of a quantum
theory in the most general sense---kinematic independence,
noncommutativity, and nonlocality---we have derived two of the three
information-theoretic constraints: the impossibility of
superluminal information transfer
between two physical systems by performing measurements on
one of them, and the impossibility of perfectly broadcasting the
information contained in an unknown physical state.

It remains an open
question whether the third information-theoretic constraint---the
impossibility of unconditionally secure bit commitment---can be
derived as well. As we indicated above, this would involve something
equivalent to an algebraic generalization of
 the Hughston-Jozsa-Wootters theorem \cite{Hughston} to cover
 cases of systems with an infinite number of degrees of freedom
  that arise in quantum field theory and the thermodynamic limit
 of quantum statistical mechanics (in which the number of
 microsystems and the volume they occupy goes to infinity, while the
 density defined by their ratio remains constant).
 The Stone-von Neumann theorem, which guarantees the existence of a
 unique representation (up to unitary equivalence) of the canonical
 commutation relations for systems with a finite number of degrees of
 freedom, breaks down for such cases, and there will
 be many unitarily inequivalent
 representations of the canonical commutation relations.

Since we
 intend our characterization of quantum theory to apply quite
 generally to these cases as well (including the quantum theoretical
 description of exotic phenomena
 such as Hawking radiation, black hole evaporation, Hawking
 information loss, etc.), we do not restrict the notion
 of a quantum theory to the
 standard quantum mechanics of a system represented on a single
 Hilbert space with a unitary dynamics.  So it would not be an
 appropriate goal of our characterization project to
 derive the Schr\"{o}dinger equation as a description of the dynamics
 of a quantum
 system from information-theoretic assumptions.
 A unitary dynamics will not be implementable
 in a quantum field theory on a curved space-time, for example, which
might be a
 preliminary semi-classical step towards a quantum theory of gravity
 (see Arageorgis \textit{et al} \cite{Ara}).

The foundational significance of our derivation, as we see it, is
that
quantum mechanics should be interpreted as a \textit{principle
theory}, where
the principles at issue are information-theoretic. The
 distinction between \textit{principle} and
\textit{constructive} theories is introduced by Einstein in
his discussion of the significance of the transition from Newtonian
to relativistic physics \cite{Einstein1}. As Einstein puts
it, most theories in physics are constructive, with the aim of
representing
complex phenomena as constructed out of the elements of a simple
formal scheme. So, for example,
the kinetic theory of gases is a constructive theory of
thermal and diffusion processes in terms of the movement of
molecules.
By contrast \cite[p. 228]{Einstein1}, principle theories begin with
empirically
discovered
`general characteristics of natural processes, principles that give
rise to mathematically formulated criteria which the separate
processes or the theoretical representations of them have to
satisfy.' Einstein cites
thermodynamics as the paradigm example of a principle theory. The
methodology here is analytic, not synthetic, with the aim of
deducing \cite[p. 228]{Einstein1}
`necessary conditions, which separate events have to satisfy, from
the
universally experienced fact that perpetual motion is impossible.'

Einstein's point is that the theory of relativity is to be understood
as a principle theory. In the case of the
special theory, there are two relevant principles:
the equivalence of inertial frames for all physical laws
(the laws of electromagnetic phenomena as well as the laws of
mechanics), and
the constancy of the velocity of light in vacuo for
all inertial frames. These
principles are irreconcilable in the Euclidean geometry of Newtonian
space-time, where inertial frames are related by Galilean
transformations. The required revision yields the special theory of
relativity and Minkowski geometry, in
which inertial frames are related by Lorentz transformations.
In his `Autobiographical Notes' \cite[p. 57]{Einstein3}, Einstein
characterizes the special principle of relativity,
that the laws of physics are invariant with
respect to Lorentz transformations from one inertial system to
another, as `a restricting principle for natural laws, comparable to
the restricting principle of the non-existence of the
\textit{perpetuum mobile} which underlies thermodynamics.' In the
case of the general theory of relativity, the group of
allowable transformations includes
 all differentiable transformations
of the space-time
manifold onto itself.

A relativistic theory is a theory with certain symmetry or invariance
properties, defined in terms of a group of space-time
transformations.
Following Einstein we understand this invariance to be a consequence
of the fact that we live in a world in which natural processes are
subject to certain constraints. A quantum theory is
a theory in which the observables and states have a certain
characteristic
algebraic structure. Unlike relativity theory, quantum mechanics was
born as a recipe or algorithm for calculating the expectation values
of observables measured by macroscopic measuring instruments. These
expectation values (or probabilities of ranges of values of
observables) cannot be reduced to probability distributions over the
values of dynamical variables (or probability distributions over
properties of the system).
Analogously, one might imagine that the special theory of relativity
was
first formulated geometrically by Minkowski rather than Einstein, as
an algorithm for relativistic kinematics and the Lorentz
transformation, which is incompatible with the kinematics of
Newtonian
space-time.
What differentiates
the two cases is that Einstein's derivation provides an
interpretation for relativity theory:
a description of the conditions under which the theory
would be true, in terms of certain principles that constrain the
law-like behavior of physical systems.
It is in this sense that our derivation of
quantum theory from information-theoretic principles can be
understood
as an
interpretation of quantum theory: the theory can now be seen as
reflecting the constraints imposed on the theoretical representations
of physical processes by these principles.

\section*{Acknowledgements}
We thank Rob Spekkens for insightful comments on broadcasting and bit
commitment that were very helpful in clearing up some confusions and
unclarities.  One of us (J.B.) wishes to acknowledge support from the
University of Maryland for a Leave Fellowship and a General Research
Board Fellowship for the duration of the project.

\end{document}